\documentclass[prb, twocolumn, superscriptaddress]{revtex4-1}
\usepackage{graphicx}% Include figure files
\usepackage{dcolumn}% Align table columns on decimal point
\usepackage{color}
\usepackage{caption}
\usepackage{subcaption}
\usepackage{soul}

\setcitestyle{square}

\begin{document}

%stuff 

%\title{Low energy BCS $d$-wave electrodynamics of an electron-doped cuprate superconductor} 
\title{BCS \textit{d}-wave behavior in the THz electrodynamic response of electron-doped cuprate superconductors} 

\author{Zhenisbek Tagay}
\affiliation{Department of Physics and Astronomy, The Johns Hopkins University, Baltimore, MD 21218, USA}
\author{Fahad Mahmood}
\affiliation{Department of Physics and Astronomy, The Johns Hopkins University, Baltimore, MD 21218, USA}
\affiliation{Department of Physics, University of Illinois at Urbana-Champaign, Urbana, 61801 IL, USA}
\affiliation{F. Seitz Materials Research Laboratory, University of Illinois at Urbana-Champaign, Urbana, 61801 IL, USA}
\author{Anaelle Legros}
\affiliation{Department of Physics and Astronomy, The Johns Hopkins University, Baltimore, MD 21218, USA}
\author{Tarapada Sarkar}
\affiliation{Department of Physics, Maryland Quantum Materials Center, University of Maryland, College Park, MD 20742, USA}
\author{Richard L. Greene}
\affiliation{Department of Physics, Maryland Quantum Materials Center, University of Maryland, College Park, MD 20742, USA}
\author{N. P. Armitage}
\affiliation{Department of Physics and Astronomy, The Johns Hopkins University, Baltimore, MD 21218, USA}
\begin{abstract}
        Although cuprate superconductors have been intensively studied for the past decades, there is no consensus regarding the  microscopic origin of their superconductivity. In this work, we measure the low-energy electrodynamic response of slightly underdoped and overdoped La$_{2-x}$Ce$_x$CuO$_4$ thin films using time-domain terahertz (THz) spectroscopy to determine the temperature and field dependence of the superfluid spectral weight. We show that the temperature dependence obeys the relation \textit{n$_s$} $\propto$ $1-(T/T_c)^2$, typical for dirty limit BCS-like $d$-wave superconductors. Furthermore, the magnetic field dependence was found to follow a sublinear $\sqrt{B}$ form, which supports predictions based on a $d$-wave symmetry for the superconducting gap. These observations imply that the superconducting order in these electron-doped cuprates can be well described in terms of a disordered BCS $d$-wave formalism.
\end{abstract}

%Recent experiments on the hold doped cuprates have shown the inconsistency of superfluid response with conventional BCS-framework, whereas other studies have reported possible pairing symmetry transition on the overdoped side of the phase diagram \textcolor{blue}{(refs?)}

\maketitle

%\tableofcontents
\section{Introduction}

The origin of superconducting order in cuprates is still debated. Recent optical and mutual inductance experiments in the hole-doped cuprate La$_{2-x}$Sr$_x$CuO$_4$ have shown the inconsistency of experimental results with aspects of the standard Bardeen-Cooper-Schrieffer (BCS) description  \cite{Nature, PhysRevLett.122.027003}. They were followed by a few theoretical works arguing that the measurements can in fact be almost entirely understood within dirty-limit BCS theory with certain assumptions\cite{PhysRevB.96.024501, PhysRevB.98.054506, PhysRevResearch.2.013228}. However, questions remain whether these assumptions are reasonable for cuprates and require further investigation.   For electron-doped cuprates, although it has been demonstrated that the temperature dependence of superfluid density is quadratic and agrees with BCS dirty $d$-wave formalism \cite{RevModPhys.82.2421,PhysRevLett.85.3700,PhysRevLett.92.157005,PhysRevLett.85.3696}, measurements of the magnetic penetration depth $\lambda$ in Pr$_{2-x}$Ce$_x$CuO$_4$ (PCCO) and La$_{2-x}$Ce$_x$CuO$_4$ (LCCO) have provided evidence for a transition to a fully gapped superconducting state on the overdoped side by showing an exponential temperature dependence of $\lambda(T)$ \cite{PhysRevLett.88.207005, PhysRevB.68.054511}. Another experimental study on PCCO has shown a weak temperature dependence of $\lambda(T)$ at low temperatures and claimed that mixed gap symmetries might be a possible explanation\cite{PhysRevLett.91.087001}. It was followed by a theoretical work arguing that weakly coupled two-band $d$-wave model could actually account for such anomaly\cite{PhysRevLett.94.027001}. Finally, mutual inductance studies in overdoped LCCO have reported a field dependence of the penetration depth which differ slightly from the $d$-wave picture\cite{pssb.200301692}. 

To further investigate this cuprate puzzle, we use time-domain terahertz spectroscopy (TDTS) to probe the low-energy electrodynamics of electron-doped LCCO thin films, both on the underdoped and overdoped sides. Conductivity data presented in this work was found to manifest most of the key features predicted for disordered $d$-wave BCS-superconductors\cite{}, including a residual Drude peak in the $T\rightarrow$0 limit and a characteristic upturn at high energies\cite{ PhysRevB.49.1397,PhysRevB.53.8575,PhysRevB.98.054506}. We also report the magnetic field and temperature dependences of the superconducting spectral weight. Our results imply that both overdoped and underdoped LCCO samples studied in this work can be rigorously described within a simple form of dirty limit $d$-wave BCS theory. 
\begin{figure*}[t!]
    \centering
    \includegraphics[width=1.6\columnwidth]{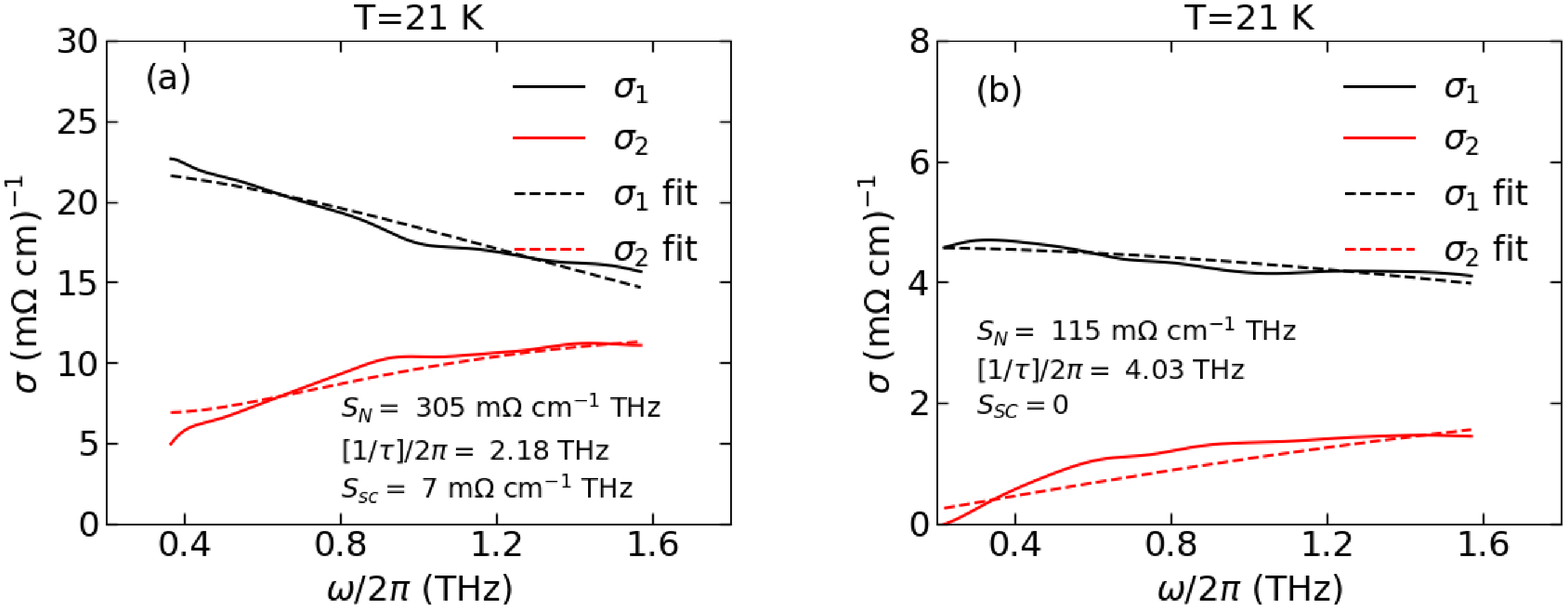}
    \caption{Drude fits to $\sigma(\omega)$ near $T_c$ (a) $x$ = 0.13, $T_c$ = 21 K (b) $x$ = 0.1, $T_c$ = 19 K}
\end{figure*}
\section{Methods}
TDTS allows one to simultaneously measure both the magnitude and phase of a THz signal. By performing TDTS measurements on both a thin film and on a reference substrate, one can obtain the complex transmission $T(\omega)$ intrinsic to the sample. Optical conductivity can then be calculated using the thin film approximation
\begin{equation}
    \sigma(\omega)=\frac{n+1}{d Z_0}\bigg[\frac{\exp[i\frac{\omega}{c} \Delta L (n-1)]}{T(\omega)}-1\bigg]
\end{equation}
where $n$ is the refractive index of the bare substrate, $Z_0$ is the impedance of free space, $d$ the thin film thickness, and $\Delta L$ the thickness difference between sample and bare substrate. $\Delta L$ is determined in two ways that gave measures with a micron of each other.  According to the Drude model, for a metal at frequencies well below the normal state scattering rate of electrons, the real part of conductivity, $\sigma_1(\omega)$, is expected to be almost frequency independent, whereas the imaginary part, $\sigma_2(\omega)$, should have very small positive value ($\sigma_2(\omega)\rightarrow0$ as $1/\tau\rightarrow\infty$). As the scattering rate is large at high temperature, we set $\Delta L$ by the assumption that these conditions were fulfilled at room temperature.  This  $\Delta L$ agreed with the value extracted by direct measure with a digital micrometer to within  1 $\mu$m.  This is sufficient for phase sensitive measurements in THz range and none of our conclusions depended on this residual uncertainty. It is also important to note that in order to get purely $ab$-plane conductivity response from this particular sample, the polarization of the incident THz signal must be properly tuned. Specific details will be addressed in Section IV.  

Here we present conductivity down to 1.6 K and in an applied field up to 3 T (the magnetic field being applied along the $c$-axis). Measurements were performed on LCCO films with $x$ = 0.13 (overdoped, $T_c=21$ K) and $x$ = 0.10 (slightly underdoped, $T_c=19$ K) compositions. Both films had thicknesses of about 75 nm and were grown using pulsed laser deposition (PLD) technique on LSAT substrates ($5\times5$ mm$^2$) at $T$ = 750$^{\circ}$C utilizing KrF excimer laser. The films were then post annealed at 600$^{\circ}$C in an oxygen at pressure of $1\times10^{-5}$  Torr for 30 minutes to induce superconductivity.  Details of the films' properties as a function of temperature and magnetic field can be found in previous work\cite{PhysRevB.97.014522,PhysRevB.96.155449,PhysRevB.98.224503}.

\section{Experimental results}
To analyze the experimental data, we consider a two-fluid model of conductivity\cite{PhysRevLett.1.399}, in which only some fraction of electrons are condensed into a superfluid and the rest remain normal. The complex conductivity then has the form $\sigma(\omega)=S_n\tau/(1-i\omega\tau)+\pi S_{sc}\delta(\omega=0)/2+i S_{sc}/\omega$, where $S_n$ and $1/\tau$ are normal state spectral weight and scattering rate and $S_{sc}$ is the superconducting spectral weight. 
Figs. 2(a) and 2(b) show the temperature dependence of the conductivity in the overdoped LCCO sample ($x$ = 0.13) with $T_c$ = 20 K at zero magnetic field. In the superconducting state ($T<T_c$), in addition to $\delta(\omega=0)$ term, the low-frequency behavior of the real part $\sigma_1(\omega)$ is dictated by impurity scattering in the form of a narrow Drude-like peak, that persists down to the lowest temperature ($T$ = 1.6 K).  At higher frequency, $\sigma_1(\omega)$ is gradually increasing. $\sigma_2(\omega)$ is fairly simple in the superconducting state at the lowest temperature and exhibits the expected $1/\omega$ dependence. In the normal state ($T$$>$$T_c$), $S_{sc}$ vanishes and all features of $\sigma_1(\omega)$ get smeared into a single broad Drude-like peak, whereas $\sigma_2(\omega)$ significantly drops in magnitude, but gradually increases with $\omega$. 
To show that Drude model works well for describing the low energy electrodynamics of LCCO samples, we perform simultaneous Drude fits to both $\sigma_1(\omega)$ and $\sigma_2(\omega)$ near $T_c$ in Figs. 1(a) and 1(b). As one can see, there is a good agreement between experimental data and theoretical fits.

%This normal state behavior is in a good agreement with metallic Drude model at frequencies below impurities scattering rate.

Several scenarios could explain the high-frequency upturn of $\sigma_1(\omega)$ in the superconducting state.  In weak-coupling $d$-wave theory, such a high-frequency upturn may be due to the presence of inelastic scatterers (\textit{e.g.} spin fluctuations\cite{ PhysRevB.53.8575,PhysRevB.98.054506}), or from the breaking of Cooper pairs in a $d$-wave superconductor\cite{PhysRevB.49.1397}. In either case, conductivity in the superconducting state is expected to reach the normal state value near frequencies of 4$\Delta$ and 8$\Delta$, respectively ($\Delta$ being the superconducting gap). Penetration depth experiment on LCCO with similar dopings ($x$ = 0.135, $T_c$ = 21.7 K) have reported $\Delta$ to be 1.25$k_BT_c$\cite{Skinta}. Using this approximate value, $\Delta$ for $x$ = 0.13 sample was found to be $0.52$ THz. We extrapolated $\sigma_1(\omega)$ to determine the frequency that the conductivity returns to the normal value. One one can see in Fig. 3(b) that such convergence happens around $2.5$ THz, or $4.8\Delta$. Although this result does not exactly match either of the earlier predictions precisely, our extrapolation is only a rough measure of the precise functional form. However, since the extrapolated result is closer to $4\Delta$ than 8$\Delta$, we infer that the inelastic scattering scenario is at least not ruled out as the the origin of the high frequency upturn in $\sigma_1(\omega)$.

\begin{figure*}[t]
    \centering
    \includegraphics[width = 1.95\columnwidth]{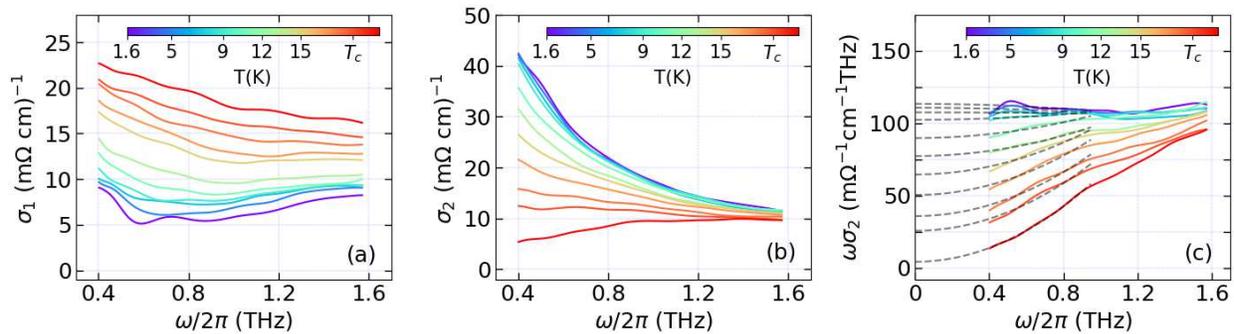}
    \caption{(a) Real and (b) imaginary parts of optical conductivity for overdoped LCCO (x=0.13) as a function of frequency and temperature. (c) $\omega\sigma_2$ derived from $\sigma_2(\omega)$ data. Dashed lines indicate linear extrapolation to $\omega$=0}
    \label{fig:Fig1}
\end{figure*}

\begin{figure*}[t]
    \centering
    \includegraphics[width = 1.95\columnwidth]{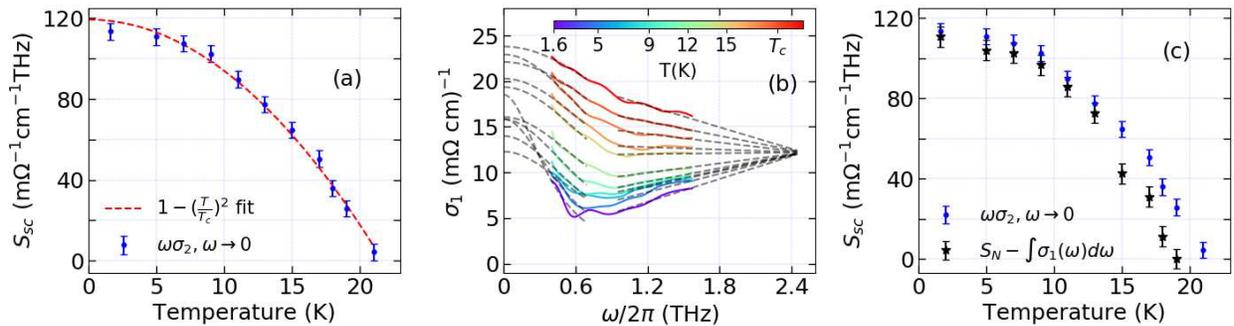}
    \caption{(a) Temperature dependence of the superfluid spectral weight $S_{sc}$ for the $x=0.13$ sample. Red dashed line denotes quadratic fit.    (b) Fitting $\sigma_1(\omega)$ to Drude (low frequency) and linear forms (high frequency).  Grey dashed lines are extrapolations as described in the text.
    (c) Comparison of $S_{sc}$ obtained using 2 different methods.   }
    \label{fig:Fig2}
\end{figure*}

\begin{figure*}
    \centering
    \includegraphics[width=1.95\columnwidth]{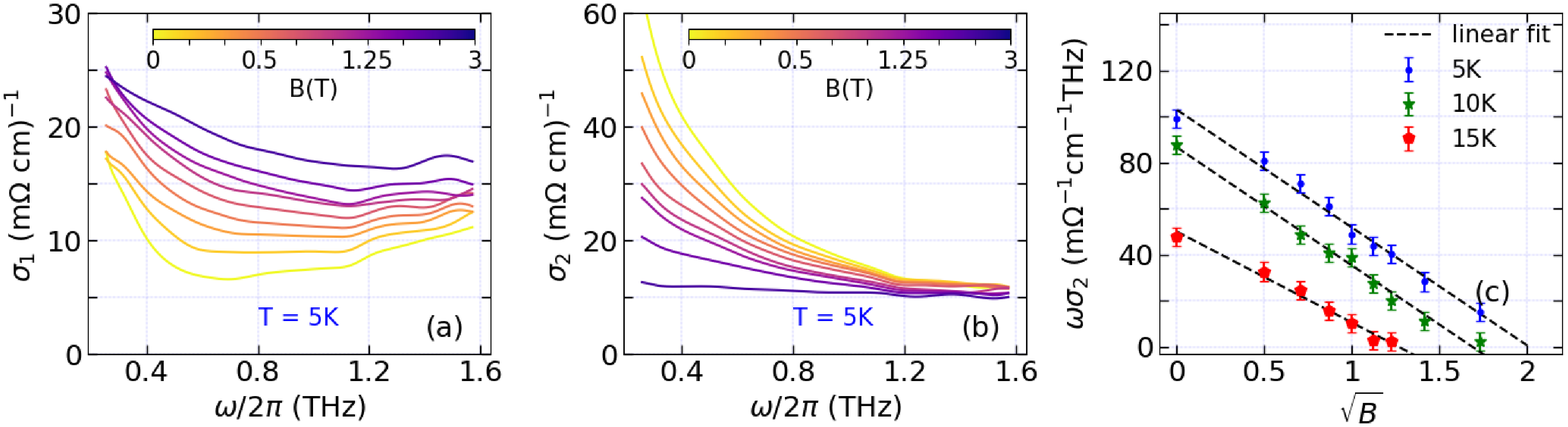}
    \caption{(a) Real and (b) imaginary parts of optical conductivity for overdoped LCCO (x=0.13) at different magnetic fields (c) Superfluid spectral weight $S_{sc}$ at different temperatures as a function of applied magnetic field. Dashed lines indicate linear fit.}
    \label{fig:Fig3}
\end{figure*}

We use $\sigma_2(\omega)$ to calculate $S_{sc}$, which is proportional to the superfluid density.  Although each electronic channel contributes to the conductivity in a different functional form, the superfluid component, $iS_{sc}/\omega$, is the only term which exhibits a $1/\omega$ dependence as $\omega \rightarrow 0$. This enables us to extract the spectral weight $S_{sc}$ by extrapolating $\omega\sigma_2(\omega)$ with a quadratic curve down to $\omega$ = 0. ($\sigma_2(\omega)\approx S_n\tau^2\omega+S_{sc}/\omega$ at low frequencies when $\omega\tau<<1$) We start with plotting $\omega\sigma_2(\omega)$ as a function of frequency for different temperatures as shown in Fig. 2(c). The data at $T$ = 1.6 K is flat, indicating that at low temperatures $\sigma_2(\omega)$ is almost fully composed of the superfluid component. As we go to higher temperatures, the slope becomes steeper, implying the reduction of $S_{sc}$. The temperature dependence of the extracted superfluid spectral weight is depicted in Fig. 3(a). The data can be fitted with good precision to a quadratic form, characteristic of dirty-limit $d$-wave superconductors with primarily unitary scattering\cite{PhysRevB.50.10250}. In fact, similar results were observed in penetration depth measurements on PCCO\cite{PhysRevLett.85.3700,PhysRevLett.92.157005}, but data showed  a $T^2$-dependence only for  $T<$ 0.3 $T_c$. Such a behavior was predicted for $d$-wave superconductors with strong scattering (unitarity limit) where the penetration depth would cross over from quadratic to linear temperature dependence at $T^{*}\simeq$ 6 ln2 $\gamma/\pi$, $\gamma$ being the width of the residual Drude peak in $\sigma_1(\omega)$ in the $T\rightarrow$ 0 limit\cite{PhysRevB.48.4219}.  Based on our experimental data,  $T^{*}$ for this overdoped film is found to be greater than $T_c$,  which may explain the absence of crossover in Fig. 3(a). 

%One can notice that at $T$ = 21 K (above $T_c$) extrapolation of $\omega\sigma_2(\omega)$ results in negative $S_{sc}$ instead of approaching zero value expected for Drude model. The possible explanation for this mismatch is a mixing of $ab$-plane and $c$-axis response functions of a sample (see Section IV) which can lead to additional non-Drude terms in measured $\sigma(\omega)$. Although tuning the polarization of incoming THz beam can minimize this effect, the $c$-axis contribution can still result in small deviation of $S_{sc}$, especially for temperatures close to $T_c$.

\begin{figure*}[]
    \includegraphics[width=1.95\columnwidth]{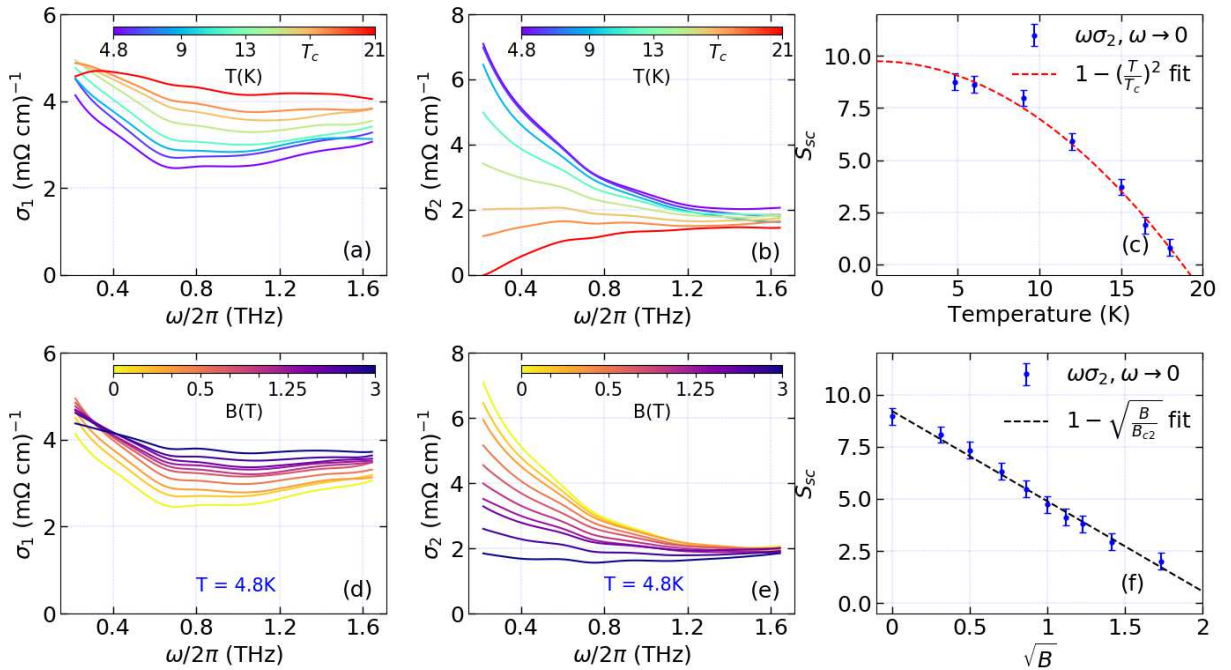}
    \caption{Complex conductivity of x=0.1 underdoped sample at: (a)-(b) different temperatures (B=0 T) , (d)-(e) different magnetic fields (T=4.8 K). (c) Temperature and (f) field dependence of extracted superconducting spectral weight $S_{sc}$. Red dashed line denotes quadratic fit to data. Blue line indicates linear fit. }
    \label{fig:Fig4}
\end{figure*}

\begin{figure*}[]
    \includegraphics[width=1.6\columnwidth]{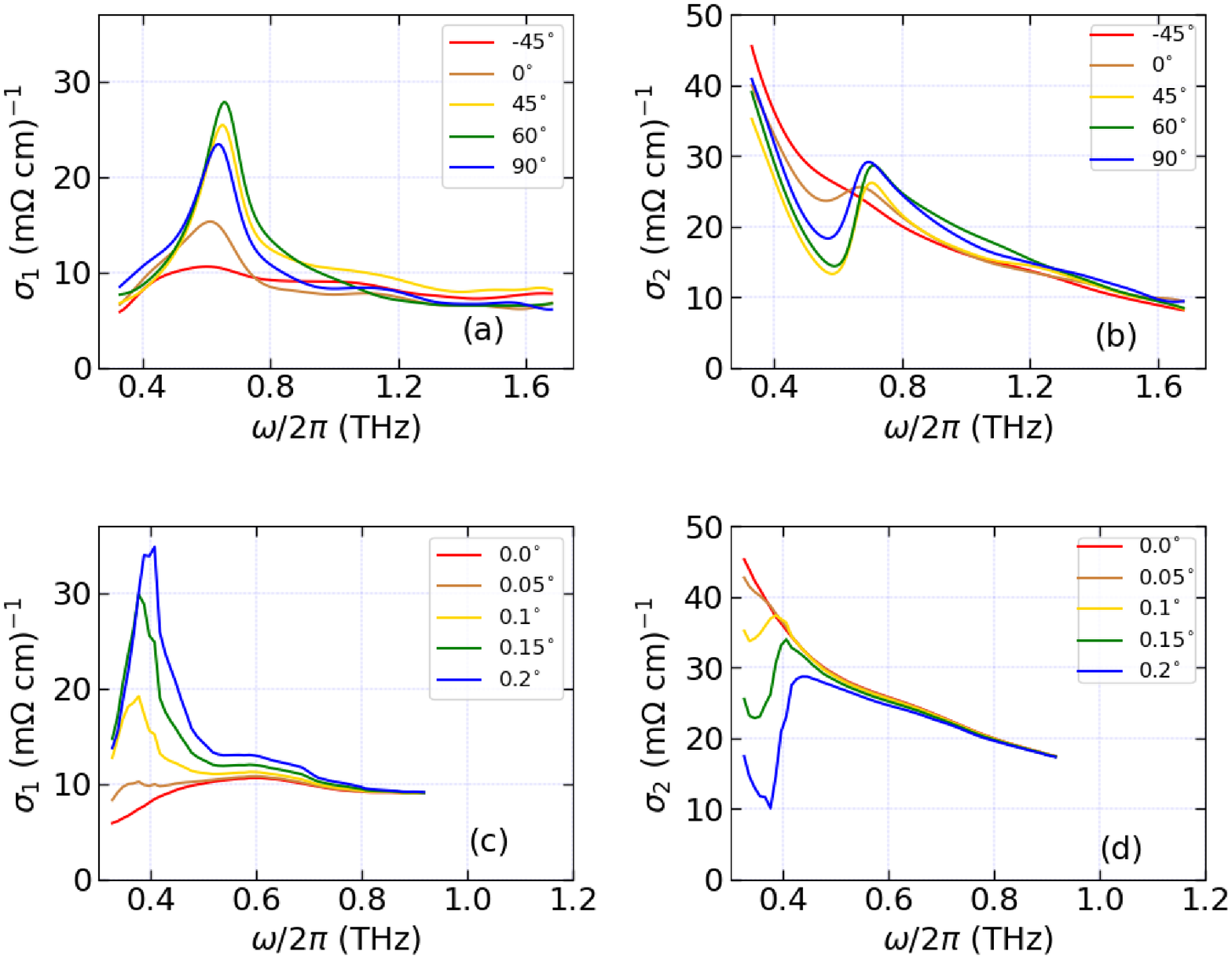}
    \caption{\textbf{(a)} and \textbf{(b)}: Real and imaginary part, respectively, of the measured in-plane conductivity of LCCO \textit{x} = 0.13 as a function of frequency, at \textit{T} = 5 K, for various polarizations of the incident THz beam (changing the angle $\Phi$ of the linear polarization). \textbf{(c)} and \textbf{(d)}: Real and imaginary part, respectively, of the calculated in-plane $\sigma_{eff}$ as a function of frequency, in the case of a contribution of the \textit{c}-axis conductivity via a tilt angle $\alpha$. The conductivity is calculated for different angles $\alpha$, using for $\sigma_{ab}$ the actual data from (a) and (b) at $\Phi$ = -45$^\circ$, and for $\sigma_c$ the extracted data at \textit{x} = 0.081 and \textit{T} = 2.5K of reference Pimenov \textit{et al.}\cite{pimenov2002peak}.}
    \label{fig:Fig5}
\end{figure*}

Another way to extract the superfluid density is by exploiting sum rules. In the context of the Ferrel-Glover-Tinkham rule \cite{PhysRevLett.2.331, PhysRev.109.1398}, the sum of the spectral weights for different states (normal and superconducting) has to be conserved. This implies that the reduction of $S_{sc}$ with temperature seen earlier should be accompanied by an enhancement in the normal state spectral weight $S_n$. Given that $S_n=\int_0^\infty\sigma_1(\omega)d\omega$, the areal difference between the real conductivities at different temperatures is expected to be equal to the associated loss in $S_{sc}$. To get a rough estimate of this, we extrapolate $\sigma_1(\omega)$ below and above our frequency range in Fig. 3(b). As discussed earlier, the low frequency part can be fitted to a Drude form $S_n\tau/(1+\omega^2\tau^2)$, with $\tau$ the relaxation time. Regarding the high frequency part, the exact functional form of an upturn in $\sigma_1(\omega)$ is sophisticated, but since all $\sigma_1(\omega)$ lines seem to converge fast, a linear fit was found to be a good estimate. In Fig. 3(c) we compare the values for superconducting spectral weight obtained this way with the ones extracted from $\omega\sigma_2(\omega)$.  One can see  that $S_{sc}$ derived via two different methods fall within the uncertainty range up to $\approx$13 K, followed by a deviation at higher temperatures. However, two independent methods result in a similar quadratic temperature dependence which is a good indication that the analysis we perform to extract $S_{sc}$ is valid. Here we want to point out that the sum rule approach is only approximate due to the limited frequency range (0.4 - 1.6 THz) of measured data and complicated form of $\sigma_1(\omega)$ at higher frequencies, and was only used to validate more precise $\omega\sigma_2$ extrapolation method. 

We then measured the dependence of the conductivity in the same overdoped sample as a function of $c$-axis magnetic field. Complex conductivity at $T$ = 5 K for fields up to 3 T are presented in Figs. 4(a)-4(b).  We use the same procedure as mentioned above to extract the superfluid spectral weight from $\omega\sigma_2(\omega)$ and then plot it as a function of applied magnetic field in Fig. 4(c). From this plot, one can see that the superfluid spectral weight reduces linearly with $\sqrt B$. This behavior, different from that of conventional $s$-wave superconductors, was first predicted by Volovik \cite{Volovik} for clean-limit $d$-wave systems. His argument was based on nonlinear London electrodynamics for gap functions with $d$-wave symmetry developed by Yip and Sauls \cite{PhysRevLett.69.2264}. A recent theoretical investigation has shown that a sublinear relation may apply to disordered materials as well\cite{PhysRevResearch.2.013228}. The Volovik effect was previously observed in optical measurements of hole-doped cuprates \cite{PhysRevLett.81.1485}, but critical fields for these compounds are usually much higher and cannot always be reached in static magnets. Data presented in Fig. 4(c) comprises a field range large enough to fully suppress superconductivity in this system and tends to exhibit $\sqrt B$ behavior throughout the entire superconducting phase. 

Several experimental studies in electron-doped cuprates, including optical\cite{PhysRevB.68.054511} and mutual inductance\cite{PhysRevLett.88.207005} measurements, have shown that the temperature dependence of $1/\lambda(T)^2$ can significantly differ for underdoped and overdoped samples. It has been stated that the superfluid response changes from $d$-wave to $s$-wave behavior at optimal doping.  The most straightforward way to investigate this is by looking at $\sigma_1$ in superconducting state.  For $s$-wave pairing symmetry the conductivity is expected to vanish below 2$\Delta$. As one can see in Fig. 2(a) and Fig. 5(a), this does not occur in the optical response at either doping level (2$\Delta\approx$ 1.04 THz for $x$ = 0.13 and $\approx$2.28 THz for $x$ = 0.1).  This rules out the possibility of pairing symmetry transition to $s$-wave.  To perform more detailed comparison of superfluid response of LCCO above and below optimal doping, the same measurements of the conductivity in temperature and magnetic field were performed on a slightly underdoped ($x$ = 0.1) sample. For this particular film,  we observe the $T_c$ to be $\approx$ 19 K which is somewhat below reported values for this doping level\cite{Greene}. We attribute this behavior to a change in oxygen content within the sample that can over a long period of time. In fact, it was shown for holed-doped cuprates that the charge carrier density can be affected by the formation of oxygen vacancies \cite{PhysRevMaterials.1.054801}. Although $T_c$-reduction would decrease the magnitude of superfluid density, the analysis in this work is purely based on how $S_{sc}$ varies with respect to certain parameters and, thus, remains valid.  

Optical conductivity for the underdoped $x=0.1$ sample at different temperatures and fields is illustrated in Fig. 5. One can see that the low-temperature electromagnetic response is qualitatively similar to that of the overdoped sample: $\sigma_1(\omega)$ exhibits a Drude-like peak at low frequencies, which is then followed by a gradual increase at higher energies. We use the same method ($\omega\sigma_2,\omega \rightarrow 0$) to extract the superfluid spectral weight. Temperature and magnetic field dependencies of the superconducting spectral weight $S_{sc}$ are presented in Fig. 5 (c) and (f), respectively.  For both dependencies, one can observe a clear match with $d$-wave functional forms mentioned above. Thereby, our experimental observations do not indicate any qualitative changes in the superconducting response crossing optimal doping.

\section{Preventing the contribution of c-axis conductivity}
%{\color{green}Experimental methods on how to remove the peak in the data, potentially due to c-axis conductivity.}\\
Even small misalignment of the thin film's c-axis can have substantial effect on the measured optical conductivity. In Figs. 6(a) and 6(b) we plot the real and imaginary parts of the conductivity for LCCO \textit{x} = 0.13 (on a different sample) at \textit{T} = 5 K ($T < T_c$) for various linear polarizations of the incident beam. An intense peak in $\sigma_1$ at finite frequency appears that becomes negligible for $\Phi$ = -45$^\circ$. This feature, only present in the superconducting state, was observed in several samples and dopings. A similar feature has been already studied in LCCO \cite{pimenov2002peak} and NCCO films \cite{pimenov2000submillimeter}. This peak has been attributed to a \textit{c}-axis leakage in the conductivitiy.  Indeed it was demonstrated that a small tilt (as small as 1$
^\circ$) of the \textit{c}-axis with respect to the sample surface's perpendicular could induce a strong contribution of the \textit{c}-axis conductivity in the measured effective in-plane conductivity $\sigma_{eff}$. This could happen if the substrate was cut close to the (001) direction but with a misfit angle $\alpha$. Therefore, a strong anisotropy could lead to a peak in $\sigma_{eff}$, that was shown to be associated with a \textit{c}-axis Josephson plasma resonance in these electron-doped cuprates.

The model for expressing the effective in-plane conductivity of a sample with a small tilt $\alpha$ of the \textit{c}-axis gives:
\begin{equation}
\sigma_{eff} = \frac{-i\epsilon_0\omega(\sigma_a cos^2(\alpha) + \sigma_c sin^2(\alpha)) + \sigma_a \sigma_c}{-i\epsilon_0\omega + \sigma_a sin^2(\alpha) + \sigma_c cos^2(\alpha)},
\end{equation}
with $\sigma_a$ and $\sigma_c$ the complex in-plane and out-of-plane conductivities respectively. To give a qualitative insight into the effect of a \textit{c}-axis leakage, we represent in Figs. 6(c) and 6(d) the real and imaginary parts respectively of a calculated $\sigma_{eff}$ for several angles $\alpha$, using for $\sigma_a$ the data of (a) and (b) at $\Phi$ = -45$^\circ$ (when the contribution of the peak is negligible) and for $\sigma_c$ the extracted data of Pimenov \textit{et al.} for LCCO x=0.081 at \textit{T} = 2.5K \cite{pimenov2002peak} (we obtain $\sigma_c$ from $\epsilon_c$ via the relation $\sigma$ = $i\epsilon_0\omega$($\epsilon_\infty$-$\epsilon$), with $\epsilon_\infty$ = 1). An intense peak indeed arises in $\sigma_1$ at finite frequency, even for small tilt angles, reminiscent of the experimental observations. These calculations and the disappearance of this intense peak in the data with a change in the linear polarization of the incoming beam seem to indicate that the feature is extrinsic and due to a mixing of \textit{c}-axis and \textit{(ab)}-plane conductivities. We chose in this work to use an angle that fully suppresses this \textit{c}-axis contribution using a specific polarization to remove these additional effects and study the pure in-plane response.
\\
\section{Conclusion}
In summary, we measured the complex optical conductivity of overdoped and slightly underdoped LCCO thin films in the THz range. Low temperature behavior in both cases was found to match theoretical predictions for $d$-wave BCS superconductors in the presence of impurities. Moreover, temperature and magnetic field dependencies of extracted superconducting spectral weights were shown to agree with extant theories. In overall, the analysis in this work strongly indicates that the low-energy electromagnetic response of this electron-doped cuprate in the vicinity of optimal doping can be well understood within dirty-limit BCS $d$-wave formalism.

\section{Acknowledgements}

We would like to acknowledge A. Pimenov and A. Pronin for helpful correspondences.   Work at JHU was supported by the NSF DMR-1905519 and at the University of Maryland by NSF  Grant DMR-2002658

\bibliography{main}

\begin{thebibliography}{32}%
\makeatletter
\providecommand \@ifxundefined [1]{%
 \@ifx{#1\undefined}
}%
\providecommand \@ifnum [1]{%
 \ifnum #1\expandafter \@firstoftwo
 \else \expandafter \@secondoftwo
 \fi
}%
\providecommand \@ifx [1]{%
 \ifx #1\expandafter \@firstoftwo
 \else \expandafter \@secondoftwo
 \fi
}%
\providecommand \natexlab [1]{#1}%
\providecommand \enquote  [1]{``#1''}%
\providecommand \bibnamefont  [1]{#1}%
\providecommand \bibfnamefont [1]{#1}%
\providecommand \citenamefont [1]{#1}%
\providecommand \href@noop [0]{\@secondoftwo}%
\providecommand \href [0]{\begingroup \@sanitize@url \@href}%
\providecommand \@href[1]{\@@startlink{#1}\@@href}%
\providecommand \@@href[1]{\endgroup#1\@@endlink}%
\providecommand \@sanitize@url [0]{\catcode `\\12\catcode `\$12\catcode
  `\&12\catcode `\#12\catcode `\^12\catcode `\_12\catcode `\%12\relax}%
\providecommand \@@startlink[1]{}%
\providecommand \@@endlink[0]{}%
\providecommand \url  [0]{\begingroup\@sanitize@url \@url }%
\providecommand \@url [1]{\endgroup\@href {#1}{\urlprefix }}%
\providecommand \urlprefix  [0]{URL }%
\providecommand \Eprint [0]{\href }%
\providecommand \doibase [0]{http://dx.doi.org/}%
\providecommand \selectlanguage [0]{\@gobble}%
\providecommand \bibinfo  [0]{\@secondoftwo}%
\providecommand \bibfield  [0]{\@secondoftwo}%
\providecommand \translation [1]{[#1]}%
\providecommand \BibitemOpen [0]{}%
\providecommand \bibitemStop [0]{}%
\providecommand \bibitemNoStop [0]{.\EOS\space}%
\providecommand \EOS [0]{\spacefactor3000\relax}%
\providecommand \BibitemShut  [1]{\csname bibitem#1\endcsname}%
\let\auto@bib@innerbib\@empty
%</preamble>
\bibitem [{\citenamefont {Božović}\ \emph {et~al.}(2016)\citenamefont
  {Božović}, \citenamefont {He}, \citenamefont {Wu},\ and\ \citenamefont
  {Bollinger}}]{Nature}%
  \BibitemOpen
  \bibfield  {author} {\bibinfo {author} {\bibfnamefont {I.}~\bibnamefont
  {Božović}}, \bibinfo {author} {\bibfnamefont {X.}~\bibnamefont {He}},
  \bibinfo {author} {\bibfnamefont {J.}~\bibnamefont {Wu}}, \ and\ \bibinfo
  {author} {\bibfnamefont {A.~T.}\ \bibnamefont {Bollinger}},\ }\href {\doibase
  10.1038/nature19061} {\bibfield  {journal} {\bibinfo  {journal} {Nature}\
  }\textbf {\bibinfo {volume} {536}},\ \bibinfo {pages} {309} (\bibinfo {year}
  {2016})}\BibitemShut {NoStop}%
\bibitem [{\citenamefont {Mahmood}\ \emph {et~al.}(2019)\citenamefont
  {Mahmood}, \citenamefont {He}, \citenamefont {Bo\ifmmode \check{z}\else
  \v{z}\fi{}ovi\ifmmode~\acute{c}\else \'{c}\fi{}},\ and\ \citenamefont
  {Armitage}}]{PhysRevLett.122.027003}%
  \BibitemOpen
  \bibfield  {author} {\bibinfo {author} {\bibfnamefont {F.}~\bibnamefont
  {Mahmood}}, \bibinfo {author} {\bibfnamefont {X.}~\bibnamefont {He}},
  \bibinfo {author} {\bibfnamefont {I.}~\bibnamefont {Bo\ifmmode \check{z}\else
  \v{z}\fi{}ovi\ifmmode~\acute{c}\else \'{c}\fi{}}}, \ and\ \bibinfo {author}
  {\bibfnamefont {N.~P.}\ \bibnamefont {Armitage}},\ }\href {\doibase
  10.1103/PhysRevLett.122.027003} {\bibfield  {journal} {\bibinfo  {journal}
  {Phys. Rev. Lett.}\ }\textbf {\bibinfo {volume} {122}},\ \bibinfo {pages}
  {027003} (\bibinfo {year} {2019})}\BibitemShut {NoStop}%
\bibitem [{\citenamefont {Lee-Hone}\ \emph {et~al.}(2017)\citenamefont
  {Lee-Hone}, \citenamefont {Dodge},\ and\ \citenamefont
  {Broun}}]{PhysRevB.96.024501}%
  \BibitemOpen
  \bibfield  {author} {\bibinfo {author} {\bibfnamefont {N.~R.}\ \bibnamefont
  {Lee-Hone}}, \bibinfo {author} {\bibfnamefont {J.~S.}\ \bibnamefont {Dodge}},
  \ and\ \bibinfo {author} {\bibfnamefont {D.~M.}\ \bibnamefont {Broun}},\
  }\href {\doibase 10.1103/PhysRevB.96.024501} {\bibfield  {journal} {\bibinfo
  {journal} {Phys. Rev. B}\ }\textbf {\bibinfo {volume} {96}},\ \bibinfo
  {pages} {024501} (\bibinfo {year} {2017})}\BibitemShut {NoStop}%
\bibitem [{\citenamefont {Lee-Hone}\ \emph {et~al.}(2018)\citenamefont
  {Lee-Hone}, \citenamefont {Mishra}, \citenamefont {Broun},\ and\
  \citenamefont {Hirschfeld}}]{PhysRevB.98.054506}%
  \BibitemOpen
  \bibfield  {author} {\bibinfo {author} {\bibfnamefont {N.~R.}\ \bibnamefont
  {Lee-Hone}}, \bibinfo {author} {\bibfnamefont {V.}~\bibnamefont {Mishra}},
  \bibinfo {author} {\bibfnamefont {D.~M.}\ \bibnamefont {Broun}}, \ and\
  \bibinfo {author} {\bibfnamefont {P.~J.}\ \bibnamefont {Hirschfeld}},\ }\href
  {\doibase 10.1103/PhysRevB.98.054506} {\bibfield  {journal} {\bibinfo
  {journal} {Phys. Rev. B}\ }\textbf {\bibinfo {volume} {98}},\ \bibinfo
  {pages} {054506} (\bibinfo {year} {2018})}\BibitemShut {NoStop}%
\bibitem [{\citenamefont {Lee-Hone}\ \emph {et~al.}(2020)\citenamefont
  {Lee-Hone}, \citenamefont {\"Ozdemir}, \citenamefont {Mishra}, \citenamefont
  {Broun},\ and\ \citenamefont {Hirschfeld}}]{PhysRevResearch.2.013228}%
  \BibitemOpen
  \bibfield  {author} {\bibinfo {author} {\bibfnamefont {N.~R.}\ \bibnamefont
  {Lee-Hone}}, \bibinfo {author} {\bibfnamefont {H.~U.}\ \bibnamefont
  {\"Ozdemir}}, \bibinfo {author} {\bibfnamefont {V.}~\bibnamefont {Mishra}},
  \bibinfo {author} {\bibfnamefont {D.~M.}\ \bibnamefont {Broun}}, \ and\
  \bibinfo {author} {\bibfnamefont {P.~J.}\ \bibnamefont {Hirschfeld}},\ }\href
  {\doibase 10.1103/PhysRevResearch.2.013228} {\bibfield  {journal} {\bibinfo
  {journal} {Phys. Rev. Research}\ }\textbf {\bibinfo {volume} {2}},\ \bibinfo
  {pages} {013228} (\bibinfo {year} {2020})}\BibitemShut {NoStop}%
\bibitem [{\citenamefont {Armitage}\ \emph {et~al.}(2010)\citenamefont
  {Armitage}, \citenamefont {Fournier},\ and\ \citenamefont
  {Greene}}]{RevModPhys.82.2421}%
  \BibitemOpen
  \bibfield  {author} {\bibinfo {author} {\bibfnamefont {N.~P.}\ \bibnamefont
  {Armitage}}, \bibinfo {author} {\bibfnamefont {P.}~\bibnamefont {Fournier}},
  \ and\ \bibinfo {author} {\bibfnamefont {R.~L.}\ \bibnamefont {Greene}},\
  }\href {\doibase 10.1103/RevModPhys.82.2421} {\bibfield  {journal} {\bibinfo
  {journal} {Rev. Mod. Phys.}\ }\textbf {\bibinfo {volume} {82}},\ \bibinfo
  {pages} {2421} (\bibinfo {year} {2010})}\BibitemShut {NoStop}%
\bibitem [{\citenamefont {Prozorov}\ \emph {et~al.}(2000)\citenamefont
  {Prozorov}, \citenamefont {Giannetta}, \citenamefont {Fournier},\ and\
  \citenamefont {Greene}}]{PhysRevLett.85.3700}%
  \BibitemOpen
  \bibfield  {author} {\bibinfo {author} {\bibfnamefont {R.}~\bibnamefont
  {Prozorov}}, \bibinfo {author} {\bibfnamefont {R.~W.}\ \bibnamefont
  {Giannetta}}, \bibinfo {author} {\bibfnamefont {P.}~\bibnamefont {Fournier}},
  \ and\ \bibinfo {author} {\bibfnamefont {R.~L.}\ \bibnamefont {Greene}},\
  }\href {\doibase 10.1103/PhysRevLett.85.3700} {\bibfield  {journal} {\bibinfo
   {journal} {Phys. Rev. Lett.}\ }\textbf {\bibinfo {volume} {85}},\ \bibinfo
  {pages} {3700} (\bibinfo {year} {2000})}\BibitemShut {NoStop}%
\bibitem [{\citenamefont {Snezhko}\ \emph {et~al.}(2004)\citenamefont
  {Snezhko}, \citenamefont {Prozorov}, \citenamefont {Lawrie}, \citenamefont
  {Giannetta}, \citenamefont {Gauthier}, \citenamefont {Renaud},\ and\
  \citenamefont {Fournier}}]{PhysRevLett.92.157005}%
  \BibitemOpen
  \bibfield  {author} {\bibinfo {author} {\bibfnamefont {A.}~\bibnamefont
  {Snezhko}}, \bibinfo {author} {\bibfnamefont {R.}~\bibnamefont {Prozorov}},
  \bibinfo {author} {\bibfnamefont {D.~D.}\ \bibnamefont {Lawrie}}, \bibinfo
  {author} {\bibfnamefont {R.~W.}\ \bibnamefont {Giannetta}}, \bibinfo {author}
  {\bibfnamefont {J.}~\bibnamefont {Gauthier}}, \bibinfo {author}
  {\bibfnamefont {J.}~\bibnamefont {Renaud}}, \ and\ \bibinfo {author}
  {\bibfnamefont {P.}~\bibnamefont {Fournier}},\ }\href {\doibase
  10.1103/PhysRevLett.92.157005} {\bibfield  {journal} {\bibinfo  {journal}
  {Phys. Rev. Lett.}\ }\textbf {\bibinfo {volume} {92}},\ \bibinfo {pages}
  {157005} (\bibinfo {year} {2004})}\BibitemShut {NoStop}%
\bibitem [{\citenamefont {Kokales}\ \emph {et~al.}(2000)\citenamefont
  {Kokales}, \citenamefont {Fournier}, \citenamefont {Mercaldo}, \citenamefont
  {Talanov}, \citenamefont {Greene},\ and\ \citenamefont
  {Anlage}}]{PhysRevLett.85.3696}%
  \BibitemOpen
  \bibfield  {author} {\bibinfo {author} {\bibfnamefont {J.~D.}\ \bibnamefont
  {Kokales}}, \bibinfo {author} {\bibfnamefont {P.}~\bibnamefont {Fournier}},
  \bibinfo {author} {\bibfnamefont {L.~V.}\ \bibnamefont {Mercaldo}}, \bibinfo
  {author} {\bibfnamefont {V.~V.}\ \bibnamefont {Talanov}}, \bibinfo {author}
  {\bibfnamefont {R.~L.}\ \bibnamefont {Greene}}, \ and\ \bibinfo {author}
  {\bibfnamefont {S.~M.}\ \bibnamefont {Anlage}},\ }\href {\doibase
  10.1103/PhysRevLett.85.3696} {\bibfield  {journal} {\bibinfo  {journal}
  {Phys. Rev. Lett.}\ }\textbf {\bibinfo {volume} {85}},\ \bibinfo {pages}
  {3696} (\bibinfo {year} {2000})}\BibitemShut {NoStop}%
\bibitem [{\citenamefont {Skinta}\ \emph {et~al.}(2002)\citenamefont {Skinta},
  \citenamefont {Kim}, \citenamefont {Lemberger}, \citenamefont {Greibe},\ and\
  \citenamefont {Naito}}]{PhysRevLett.88.207005}%
  \BibitemOpen
  \bibfield  {author} {\bibinfo {author} {\bibfnamefont {J.~A.}\ \bibnamefont
  {Skinta}}, \bibinfo {author} {\bibfnamefont {M.-S.}\ \bibnamefont {Kim}},
  \bibinfo {author} {\bibfnamefont {T.~R.}\ \bibnamefont {Lemberger}}, \bibinfo
  {author} {\bibfnamefont {T.}~\bibnamefont {Greibe}}, \ and\ \bibinfo {author}
  {\bibfnamefont {M.}~\bibnamefont {Naito}},\ }\href {\doibase
  10.1103/PhysRevLett.88.207005} {\bibfield  {journal} {\bibinfo  {journal}
  {Phys. Rev. Lett.}\ }\textbf {\bibinfo {volume} {88}},\ \bibinfo {pages}
  {207005} (\bibinfo {year} {2002})}\BibitemShut {NoStop}%
\bibitem [{\citenamefont {Pronin}\ \emph {et~al.}(2003)\citenamefont {Pronin},
  \citenamefont {Pimenov}, \citenamefont {Loidl}, \citenamefont {Tsukada},\
  and\ \citenamefont {Naito}}]{PhysRevB.68.054511}%
  \BibitemOpen
  \bibfield  {author} {\bibinfo {author} {\bibfnamefont {A.~V.}\ \bibnamefont
  {Pronin}}, \bibinfo {author} {\bibfnamefont {A.}~\bibnamefont {Pimenov}},
  \bibinfo {author} {\bibfnamefont {A.}~\bibnamefont {Loidl}}, \bibinfo
  {author} {\bibfnamefont {A.}~\bibnamefont {Tsukada}}, \ and\ \bibinfo
  {author} {\bibfnamefont {M.}~\bibnamefont {Naito}},\ }\href {\doibase
  10.1103/PhysRevB.68.054511} {\bibfield  {journal} {\bibinfo  {journal} {Phys.
  Rev. B}\ }\textbf {\bibinfo {volume} {68}},\ \bibinfo {pages} {054511}
  (\bibinfo {year} {2003})}\BibitemShut {NoStop}%
\bibitem [{\citenamefont {Kim}\ \emph {et~al.}(2003)\citenamefont {Kim},
  \citenamefont {Skinta}, \citenamefont {Lemberger}, \citenamefont {Tsukada},\
  and\ \citenamefont {Naito}}]{PhysRevLett.91.087001}%
  \BibitemOpen
  \bibfield  {author} {\bibinfo {author} {\bibfnamefont {M.-S.}\ \bibnamefont
  {Kim}}, \bibinfo {author} {\bibfnamefont {J.~A.}\ \bibnamefont {Skinta}},
  \bibinfo {author} {\bibfnamefont {T.~R.}\ \bibnamefont {Lemberger}}, \bibinfo
  {author} {\bibfnamefont {A.}~\bibnamefont {Tsukada}}, \ and\ \bibinfo
  {author} {\bibfnamefont {M.}~\bibnamefont {Naito}},\ }\href {\doibase
  10.1103/PhysRevLett.91.087001} {\bibfield  {journal} {\bibinfo  {journal}
  {Phys. Rev. Lett.}\ }\textbf {\bibinfo {volume} {91}},\ \bibinfo {pages}
  {087001} (\bibinfo {year} {2003})}\BibitemShut {NoStop}%
\bibitem [{\citenamefont {Luo}\ and\ \citenamefont
  {Xiang}(2005)}]{PhysRevLett.94.027001}%
  \BibitemOpen
  \bibfield  {author} {\bibinfo {author} {\bibfnamefont {H.~G.}\ \bibnamefont
  {Luo}}\ and\ \bibinfo {author} {\bibfnamefont {T.}~\bibnamefont {Xiang}},\
  }\href {\doibase 10.1103/PhysRevLett.94.027001} {\bibfield  {journal}
  {\bibinfo  {journal} {Phys. Rev. Lett.}\ }\textbf {\bibinfo {volume} {94}},\
  \bibinfo {pages} {027001} (\bibinfo {year} {2005})}\BibitemShut {NoStop}%
\bibitem [{\citenamefont {Zuev}\ \emph {et~al.}(2003)\citenamefont {Zuev},
  \citenamefont {Lemberger}, \citenamefont {Skinta}, \citenamefont {Greibe},\
  and\ \citenamefont {Naito}}]{pssb.200301692}%
  \BibitemOpen
  \bibfield  {author} {\bibinfo {author} {\bibfnamefont {Y.}~\bibnamefont
  {Zuev}}, \bibinfo {author} {\bibfnamefont {T.~R.}\ \bibnamefont {Lemberger}},
  \bibinfo {author} {\bibfnamefont {J.~A.}\ \bibnamefont {Skinta}}, \bibinfo
  {author} {\bibfnamefont {T.}~\bibnamefont {Greibe}}, \ and\ \bibinfo {author}
  {\bibfnamefont {M.}~\bibnamefont {Naito}},\ }\href {\doibase
  10.1002/pssb.200301692} {\bibfield  {journal} {\bibinfo  {journal} {physica
  status solidi (b)}\ }\textbf {\bibinfo {volume} {236}},\ \bibinfo {pages}
  {412} (\bibinfo {year} {2003})}\BibitemShut {NoStop}%
\bibitem [{\citenamefont {Won}\ and\ \citenamefont
  {Maki}(1994)}]{PhysRevB.49.1397}%
  \BibitemOpen
  \bibfield  {author} {\bibinfo {author} {\bibfnamefont {H.}~\bibnamefont
  {Won}}\ and\ \bibinfo {author} {\bibfnamefont {K.}~\bibnamefont {Maki}},\
  }\href {\doibase 10.1103/PhysRevB.49.1397} {\bibfield  {journal} {\bibinfo
  {journal} {Phys. Rev. B}\ }\textbf {\bibinfo {volume} {49}},\ \bibinfo
  {pages} {1397} (\bibinfo {year} {1994})}\BibitemShut {NoStop}%
\bibitem [{\citenamefont {Quinlan}\ \emph {et~al.}(1996)\citenamefont
  {Quinlan}, \citenamefont {Hirschfeld},\ and\ \citenamefont
  {Scalapino}}]{PhysRevB.53.8575}%
  \BibitemOpen
  \bibfield  {author} {\bibinfo {author} {\bibfnamefont {S.~M.}\ \bibnamefont
  {Quinlan}}, \bibinfo {author} {\bibfnamefont {P.~J.}\ \bibnamefont
  {Hirschfeld}}, \ and\ \bibinfo {author} {\bibfnamefont {D.~J.}\ \bibnamefont
  {Scalapino}},\ }\href {\doibase 10.1103/PhysRevB.53.8575} {\bibfield
  {journal} {\bibinfo  {journal} {Phys. Rev. B}\ }\textbf {\bibinfo {volume}
  {53}},\ \bibinfo {pages} {8575} (\bibinfo {year} {1996})}\BibitemShut
  {NoStop}%
\bibitem [{\citenamefont {Mandal}\ \emph {et~al.}(2018)\citenamefont {Mandal},
  \citenamefont {Sarkar}, \citenamefont {Higgins},\ and\ \citenamefont
  {Greene}}]{PhysRevB.97.014522}%
  \BibitemOpen
  \bibfield  {author} {\bibinfo {author} {\bibfnamefont {P.~R.}\ \bibnamefont
  {Mandal}}, \bibinfo {author} {\bibfnamefont {T.}~\bibnamefont {Sarkar}},
  \bibinfo {author} {\bibfnamefont {J.~S.}\ \bibnamefont {Higgins}}, \ and\
  \bibinfo {author} {\bibfnamefont {R.~L.}\ \bibnamefont {Greene}},\ }\href
  {\doibase 10.1103/PhysRevB.97.014522} {\bibfield  {journal} {\bibinfo
  {journal} {Phys. Rev. B}\ }\textbf {\bibinfo {volume} {97}},\ \bibinfo
  {pages} {014522} (\bibinfo {year} {2018})}\BibitemShut {NoStop}%
\bibitem [{\citenamefont {Sarkar}\ \emph {et~al.}(2017)\citenamefont {Sarkar},
  \citenamefont {Mandal}, \citenamefont {Higgins}, \citenamefont {Zhao},
  \citenamefont {Yu}, \citenamefont {Jin},\ and\ \citenamefont
  {Greene}}]{PhysRevB.96.155449}%
  \BibitemOpen
  \bibfield  {author} {\bibinfo {author} {\bibfnamefont {T.}~\bibnamefont
  {Sarkar}}, \bibinfo {author} {\bibfnamefont {P.~R.}\ \bibnamefont {Mandal}},
  \bibinfo {author} {\bibfnamefont {J.~S.}\ \bibnamefont {Higgins}}, \bibinfo
  {author} {\bibfnamefont {Y.}~\bibnamefont {Zhao}}, \bibinfo {author}
  {\bibfnamefont {H.}~\bibnamefont {Yu}}, \bibinfo {author} {\bibfnamefont
  {K.}~\bibnamefont {Jin}}, \ and\ \bibinfo {author} {\bibfnamefont {R.~L.}\
  \bibnamefont {Greene}},\ }\href {\doibase 10.1103/PhysRevB.96.155449}
  {\bibfield  {journal} {\bibinfo  {journal} {Phys. Rev. B}\ }\textbf {\bibinfo
  {volume} {96}},\ \bibinfo {pages} {155449} (\bibinfo {year}
  {2017})}\BibitemShut {NoStop}%
\bibitem [{\citenamefont {Sarkar}\ \emph {et~al.}(2018)\citenamefont {Sarkar},
  \citenamefont {Greene},\ and\ \citenamefont
  {Das~Sarma}}]{PhysRevB.98.224503}%
  \BibitemOpen
  \bibfield  {author} {\bibinfo {author} {\bibfnamefont {T.}~\bibnamefont
  {Sarkar}}, \bibinfo {author} {\bibfnamefont {R.~L.}\ \bibnamefont {Greene}},
  \ and\ \bibinfo {author} {\bibfnamefont {S.}~\bibnamefont {Das~Sarma}},\
  }\href {\doibase 10.1103/PhysRevB.98.224503} {\bibfield  {journal} {\bibinfo
  {journal} {Phys. Rev. B}\ }\textbf {\bibinfo {volume} {98}},\ \bibinfo
  {pages} {224503} (\bibinfo {year} {2018})}\BibitemShut {NoStop}%
\bibitem [{\citenamefont {Bardeen}(1958)}]{PhysRevLett.1.399}%
  \BibitemOpen
  \bibfield  {author} {\bibinfo {author} {\bibfnamefont {J.}~\bibnamefont
  {Bardeen}},\ }\href {\doibase 10.1103/PhysRevLett.1.399} {\bibfield
  {journal} {\bibinfo  {journal} {Phys. Rev. Lett.}\ }\textbf {\bibinfo
  {volume} {1}},\ \bibinfo {pages} {399} (\bibinfo {year} {1958})}\BibitemShut
  {NoStop}%
\bibitem [{\citenamefont {Skinta}\ \emph {et~al.}(2003)\citenamefont {Skinta},
  \citenamefont {Kim}, \citenamefont {Lemberger}, \citenamefont {Greibe},\ and\
  \citenamefont {Naito}}]{Skinta}%
  \BibitemOpen
  \bibfield  {author} {\bibinfo {author} {\bibfnamefont {J.~A.}\ \bibnamefont
  {Skinta}}, \bibinfo {author} {\bibfnamefont {M.-S.}\ \bibnamefont {Kim}},
  \bibinfo {author} {\bibfnamefont {T.~R.}\ \bibnamefont {Lemberger}}, \bibinfo
  {author} {\bibfnamefont {T.}~\bibnamefont {Greibe}}, \ and\ \bibinfo {author}
  {\bibfnamefont {M.}~\bibnamefont {Naito}},\ }\href
  {https://link.springer.com/article/10.1023/A:1022962110978} {\bibfield
  {journal} {\bibinfo  {journal} {Journal of Low Temperature Physics}\ }\textbf
  {\bibinfo {volume} {131}},\ \bibinfo {pages} {359} (\bibinfo {year}
  {2003})}\BibitemShut {NoStop}%
\bibitem [{\citenamefont {Hirschfeld}\ \emph {et~al.}(1994)\citenamefont
  {Hirschfeld}, \citenamefont {Putikka},\ and\ \citenamefont
  {Scalapino}}]{PhysRevB.50.10250}%
  \BibitemOpen
  \bibfield  {author} {\bibinfo {author} {\bibfnamefont {P.~J.}\ \bibnamefont
  {Hirschfeld}}, \bibinfo {author} {\bibfnamefont {W.~O.}\ \bibnamefont
  {Putikka}}, \ and\ \bibinfo {author} {\bibfnamefont {D.~J.}\ \bibnamefont
  {Scalapino}},\ }\href {\doibase 10.1103/PhysRevB.50.10250} {\bibfield
  {journal} {\bibinfo  {journal} {Phys. Rev. B}\ }\textbf {\bibinfo {volume}
  {50}},\ \bibinfo {pages} {10250} (\bibinfo {year} {1994})}\BibitemShut
  {NoStop}%
\bibitem [{\citenamefont {Hirschfeld}\ and\ \citenamefont
  {Goldenfeld}(1993)}]{PhysRevB.48.4219}%
  \BibitemOpen
  \bibfield  {author} {\bibinfo {author} {\bibfnamefont {P.~J.}\ \bibnamefont
  {Hirschfeld}}\ and\ \bibinfo {author} {\bibfnamefont {N.}~\bibnamefont
  {Goldenfeld}},\ }\href {\doibase 10.1103/PhysRevB.48.4219} {\bibfield
  {journal} {\bibinfo  {journal} {Phys. Rev. B}\ }\textbf {\bibinfo {volume}
  {48}},\ \bibinfo {pages} {4219} (\bibinfo {year} {1993})}\BibitemShut
  {NoStop}%
\bibitem [{\citenamefont {Pimenov}\ \emph {et~al.}(2002)\citenamefont
  {Pimenov}, \citenamefont {Pronin}, \citenamefont {Loidl}, \citenamefont
  {Tsukada},\ and\ \citenamefont {Naito}}]{pimenov2002peak}%
  \BibitemOpen
  \bibfield  {author} {\bibinfo {author} {\bibfnamefont {A.}~\bibnamefont
  {Pimenov}}, \bibinfo {author} {\bibfnamefont {A.~V.}\ \bibnamefont {Pronin}},
  \bibinfo {author} {\bibfnamefont {A.}~\bibnamefont {Loidl}}, \bibinfo
  {author} {\bibfnamefont {A.}~\bibnamefont {Tsukada}}, \ and\ \bibinfo
  {author} {\bibfnamefont {M.}~\bibnamefont {Naito}},\ }\href@noop {}
  {\bibfield  {journal} {\bibinfo  {journal} {Physical Review B}\ }\textbf
  {\bibinfo {volume} {66}},\ \bibinfo {pages} {212508} (\bibinfo {year}
  {2002})}\BibitemShut {NoStop}%
\bibitem [{\citenamefont {Tinkham}\ and\ \citenamefont
  {Ferrell}(1959)}]{PhysRevLett.2.331}%
  \BibitemOpen
  \bibfield  {author} {\bibinfo {author} {\bibfnamefont {M.}~\bibnamefont
  {Tinkham}}\ and\ \bibinfo {author} {\bibfnamefont {R.~A.}\ \bibnamefont
  {Ferrell}},\ }\href {\doibase 10.1103/PhysRevLett.2.331} {\bibfield
  {journal} {\bibinfo  {journal} {Phys. Rev. Lett.}\ }\textbf {\bibinfo
  {volume} {2}},\ \bibinfo {pages} {331} (\bibinfo {year} {1959})}\BibitemShut
  {NoStop}%
\bibitem [{\citenamefont {Ferrell}\ and\ \citenamefont
  {Glover}(1958)}]{PhysRev.109.1398}%
  \BibitemOpen
  \bibfield  {author} {\bibinfo {author} {\bibfnamefont {R.~A.}\ \bibnamefont
  {Ferrell}}\ and\ \bibinfo {author} {\bibfnamefont {R.~E.}\ \bibnamefont
  {Glover}},\ }\href {\doibase 10.1103/PhysRev.109.1398} {\bibfield  {journal}
  {\bibinfo  {journal} {Phys. Rev.}\ }\textbf {\bibinfo {volume} {109}},\
  \bibinfo {pages} {1398} (\bibinfo {year} {1958})}\BibitemShut {NoStop}%
\bibitem [{\citenamefont {Volovik}(1993)}]{Volovik}%
  \BibitemOpen
  \bibfield  {author} {\bibinfo {author} {\bibfnamefont {G.~E.}\ \bibnamefont
  {Volovik}},\ }\href@noop {} {\bibfield  {journal} {\bibinfo  {journal} {JETP
  Letters}\ }\textbf {\bibinfo {volume} {58}},\ \bibinfo {pages} {469}
  (\bibinfo {year} {1993})}\BibitemShut {NoStop}%
\bibitem [{\citenamefont {Yip}\ and\ \citenamefont
  {Sauls}(1992)}]{PhysRevLett.69.2264}%
  \BibitemOpen
  \bibfield  {author} {\bibinfo {author} {\bibfnamefont {S.~K.}\ \bibnamefont
  {Yip}}\ and\ \bibinfo {author} {\bibfnamefont {J.~A.}\ \bibnamefont
  {Sauls}},\ }\href {\doibase 10.1103/PhysRevLett.69.2264} {\bibfield
  {journal} {\bibinfo  {journal} {Phys. Rev. Lett.}\ }\textbf {\bibinfo
  {volume} {69}},\ \bibinfo {pages} {2264} (\bibinfo {year}
  {1992})}\BibitemShut {NoStop}%
\bibitem [{\citenamefont {Mallozzi}\ \emph {et~al.}(1998)\citenamefont
  {Mallozzi}, \citenamefont {Orenstein}, \citenamefont {Eckstein},\ and\
  \citenamefont {Bozovic}}]{PhysRevLett.81.1485}%
  \BibitemOpen
  \bibfield  {author} {\bibinfo {author} {\bibfnamefont {R.}~\bibnamefont
  {Mallozzi}}, \bibinfo {author} {\bibfnamefont {J.}~\bibnamefont {Orenstein}},
  \bibinfo {author} {\bibfnamefont {J.~N.}\ \bibnamefont {Eckstein}}, \ and\
  \bibinfo {author} {\bibfnamefont {I.}~\bibnamefont {Bozovic}},\ }\href
  {\doibase 10.1103/PhysRevLett.81.1485} {\bibfield  {journal} {\bibinfo
  {journal} {Phys. Rev. Lett.}\ }\textbf {\bibinfo {volume} {81}},\ \bibinfo
  {pages} {1485} (\bibinfo {year} {1998})}\BibitemShut {NoStop}%
\bibitem [{\citenamefont {Jin}\ \emph {et~al.}(2011)\citenamefont {Jin},
  \citenamefont {Butch}, \citenamefont {Kirshenbaum}, \citenamefont
  {Paglione},\ and\ \citenamefont {Greene}}]{Greene}%
  \BibitemOpen
  \bibfield  {author} {\bibinfo {author} {\bibfnamefont {K.}~\bibnamefont
  {Jin}}, \bibinfo {author} {\bibfnamefont {N.}~\bibnamefont {Butch}}, \bibinfo
  {author} {\bibfnamefont {K.}~\bibnamefont {Kirshenbaum}}, \bibinfo {author}
  {\bibfnamefont {J.}~\bibnamefont {Paglione}}, \ and\ \bibinfo {author}
  {\bibfnamefont {R.~L.}\ \bibnamefont {Greene}},\ }\href {\doibase
  10.1038/nature10308} {\bibfield  {journal} {\bibinfo  {journal} {Nature}\
  }\textbf {\bibinfo {volume} {476}},\ \bibinfo {pages} {73} (\bibinfo {year}
  {2011})}\BibitemShut {NoStop}%
\bibitem [{\citenamefont {Kim}\ \emph {et~al.}(2017)\citenamefont {Kim},
  \citenamefont {Christiani}, \citenamefont {Logvenov}, \citenamefont {Choi},
  \citenamefont {Kim}, \citenamefont {Minola},\ and\ \citenamefont
  {Keimer}}]{PhysRevMaterials.1.054801}%
  \BibitemOpen
  \bibfield  {author} {\bibinfo {author} {\bibfnamefont {G.}~\bibnamefont
  {Kim}}, \bibinfo {author} {\bibfnamefont {G.}~\bibnamefont {Christiani}},
  \bibinfo {author} {\bibfnamefont {G.}~\bibnamefont {Logvenov}}, \bibinfo
  {author} {\bibfnamefont {S.}~\bibnamefont {Choi}}, \bibinfo {author}
  {\bibfnamefont {H.-H.}\ \bibnamefont {Kim}}, \bibinfo {author} {\bibfnamefont
  {M.}~\bibnamefont {Minola}}, \ and\ \bibinfo {author} {\bibfnamefont
  {B.}~\bibnamefont {Keimer}},\ }\href {\doibase
  10.1103/PhysRevMaterials.1.054801} {\bibfield  {journal} {\bibinfo  {journal}
  {Phys. Rev. Materials}\ }\textbf {\bibinfo {volume} {1}},\ \bibinfo {pages}
  {054801} (\bibinfo {year} {2017})}\BibitemShut {NoStop}%
\bibitem [{\citenamefont {Pimenov}\ \emph {et~al.}(2000)\citenamefont
  {Pimenov}, \citenamefont {Pronin}, \citenamefont {Loidl}, \citenamefont
  {Kampf}, \citenamefont {Krasnosvobodtsev},\ and\ \citenamefont
  {Nozdrin}}]{pimenov2000submillimeter}%
  \BibitemOpen
  \bibfield  {author} {\bibinfo {author} {\bibfnamefont {A.}~\bibnamefont
  {Pimenov}}, \bibinfo {author} {\bibfnamefont {A.}~\bibnamefont {Pronin}},
  \bibinfo {author} {\bibfnamefont {A.}~\bibnamefont {Loidl}}, \bibinfo
  {author} {\bibfnamefont {A.~P.}\ \bibnamefont {Kampf}}, \bibinfo {author}
  {\bibfnamefont {S.}~\bibnamefont {Krasnosvobodtsev}}, \ and\ \bibinfo
  {author} {\bibfnamefont {V.}~\bibnamefont {Nozdrin}},\ }\href@noop {}
  {\bibfield  {journal} {\bibinfo  {journal} {Applied Physics Letters}\
  }\textbf {\bibinfo {volume} {77}},\ \bibinfo {pages} {429} (\bibinfo {year}
  {2000})}\BibitemShut {NoStop}%
\end{thebibliography}%

\end{document}